\documentclass[
    ,draft            
  ]
  {aipproc}

\layoutstyle{8x11single}

\begin{document}

\newcommand{\GeV}{GeV$^2$}
\newcommand{\deut}{$^2$H}
\newcommand{\pdeut}{$\stackrel{\rightarrow}{^2\rm{H}}$}
\newcommand{\he}{$^3\rm{He}$}
\newcommand{\phe}{$\stackrel{\rightarrow}{^3\rm{He}}$}
\newcommand{\pDeen}{\pdeut($\vec{e},e^\prime n$)}
\newcommand{\Deepn}{\deut($\vec{e},e^\prime \vec{n})$}
\newcommand{\Heen}{\phe($\vec{e},e^\prime n$)}
\newcommand{\Hee}{\phe($\vec{e},e^\prime$)}
\newcommand{\Ee}{\ensuremath{E_e}} 
\newcommand{\thetae}{\ensuremath{\theta_e}} 
\newcommand{\GD}{\ensuremath{G_D}}
\newcommand{\GE}{\ensuremath{G_E}}
\newcommand{\GEp}{\ensuremath{G_E^{p}}}
\newcommand{\GEpGMp}{\ensuremath{G_E^p/G_M^p}}
\newcommand{\GEn}{\ensuremath{G_E^{n}}}
\newcommand{\GM}{\ensuremath{G_M}}
\newcommand{\GMp}{\ensuremath{G_M^{p}}}
\newcommand{\GMn}{\ensuremath{G_M^{n}{}}}
\newcommand{\Q}{\ensuremath{Q^{2}{}}}
\newcommand{\updeg}{$^{o}$}

\def\Journal#1#2#3#4{{#1} {\bf #2}, (#4) #3}
\def\NCA{Nuovo Cimento}
\def\NIM{Nucl. Instrum. Methods}
\def\NIMA{{Nucl. Instrum. Methods} A}
\def\EPJ{{Eur. Phys. Jour.} A}
\def\JPG{J. Phys. G: Nucl. Part. Phys.}
\def\NPA{{Nucl. Phys.} A}
\def\NPB{{Nucl. Phys.} B}
\def\PLB{{Phys. Lett.}  B}
\def\PRL{Phys. Rev. Lett.}
\def\PRC{{Phys. Rev.}  C}
\def\PRD{{Phys. Rev.} D}
\def\RMP{Rev. Mod. Phys.}
\def\ZPA{{Z. Phys.} A}
\def\ZPC{{Z. Phys.} C}

\title{Nucleon Form Factor Experiments and the Pion Cloud}


\author{Kees de Jager}{
  address={Thomas Jefferson National Accelerator Facility, Newport News, Virginia 23606}
}

\begin{abstract}
The experimental and theoretical status 
of elastic electron scattering from the nucleon is reviewed.  A wealth of new data of unprecedented precision, especially at small values of the momentum transfer, in parallel to new theoretical insights, has allowed sensitive tests of the influence of the pionic cloud surrounding the nucleon. 
\end{abstract}

\maketitle


\section{Introduction}
\label{intro}

The pion is considered one of the main transmitters of the force between nucleons. As a consequence, nucleons are visualized in a simplified picture to be surrounded by a cloud of pions; thus, the wave function of a proton (neutron) is expected to have a component contributed by a neutron (proton) surrounded by a positive (negative) pion. This pion cloud will then manifest itself as an extension to the charge distribution of protons and neutrons, which should be observable in the electromagnetic form factors of the nucleon at relatively small values of the momentum transfer. Nucleon electro-magnetic form factors (EMFFs) are optimally studied through the exchange of a virtual photon, in elastic electron-nucleon scattering. Polarization instrumentation,
polarized beams and targets, and the measurement of the recoil polarization have been
essential in the accurate separation of the charge and magnetic form factors and in studies
of the neutron charge form factor.

\setlength\parindent{0.25in}
Through the mid-1990s practically all available 
proton EMFF data had been collected using the Rosenbluth separation 
technique, in which the cross section is measured at fixed \Q~as a function of the linear polarization of the virtual photon $\epsilon$. Because the \GMp\ contribution to the elastic cross section
 is weighted with \Q, data on \GEp\ suffer 
from increasing systematic uncertainties with increasing \Q-values.

More than 40 years ago Akhiezer {\it et al.}\cite{akhi} (followed 20 years later by Arnold 
{\it et al.}\cite{arnold}) showed that the accuracy of nucleon charge  form-factor measurements 
could be increased significantly by scattering 
polarized electrons off a polarized target (or equivalently by
measuring the polarization of the recoiling proton). 
However, it took several decades before technology had sufficiently advanced
to make the first of such measurements feasible and only in the past few years
has a large number of new data with a significantly improved accuracy become available.
For \GEp\ measurements the highest figure of merit 
at \Q-values larger than a few \GeV\ is obtained with a focal plane polarimeter.
Here, the Jacobian focusing of the recoiling proton kinematics allows one to couple
a standard magnetic spectrometer for the proton detection to a large-acceptance
non-magnetic detector for the detection of the scattered electron. For studies of
\GEn\ one needs to use a magnetic spectrometer to detect
the scattered electron in order to cleanly identify the reaction channel. As a consequence,
the figure of merit of a polarized \phe\ target is comparable to that of a neutron polarimeter. 

\section{Proton Electric Form Factor}

In elastic electron-proton scattering a longitudinally polarized electron will transfer its 
polarization to the recoil proton. In the one-photon exchange approximation the proton 
can attain only polarization components in the scattering plane, parallel ($P_l$) and 
transverse ($P_t$) to its momentum. The ratio of the charge and magnetic form factors is
directly proportional to the ratio of these polarization components.

The greatest impact of the polarization-transfer technique was made by the two recent
experiments\cite{punj,gayou} in Hall A at Jefferson Lab, which measured the ratio \GEpGMp\ in 
a \Q-range from 0.5 to 5.6 \GeV. 
The 
most striking feature of the data is the sharp, practically linear decline 
as \Q\ increases. Since it is known that \GMp\ closely follows the dipole 
parametrization $G_{D}$, it follows that \GEp\ falls more rapidly with \Q\ 
than $G_{D}$. This significant fall-off of the form-factor ratio is in clear 
disagreement with the results from the Rosenbluth extraction.
Qattan {\it et al.}\cite{segel} performed a high-precision Rosenbluth extraction 
in Hall A at Jefferson Lab, designed specifically
to significantly reduce the systematic errors compared to earlier Rosenbluth
measurements. The main improvement came from detecting the recoiling
protons instead of the scattered electrons. One of the spectrometers was used as a 
luminosity monitor during an $\epsilon$ scan. The results\cite{segel} of this 
experiment, covering \Q-values from 2.6 to 4.1 \GeV, are in excellent agreement
with previous Rosenbluth results. This basically rules out the possibility that
the disagreement between Rosenbluth and polarization-transfer measurements
of the ratio \GEpGMp\ is due to an underestimate of $\epsilon$-dependent
uncertainties in the Rosenbluth measurements.
At the Bates Large Acceptance Spectrometer Toroid facility (BLAST, http://blast.lns.mit.edu/) at MIT a polarized hydrogen target internal to a storage ring has been used successfully to  provide highly accurate 
data on \GEp\ in a \Q-range from 0.1 to 0.6 \GeV \cite{crawford}. 

\subsection{Two-Photon Exchange}

Two-(or more-)photon exchange (TPE)  contributions to elastic electron scattering have been investigated both experimentally and theoretically for the past fifty
years. Almost all analyses
with the Rosenbluth technique have used radiative corrections that only include the infrared divergent parts of the box diagram
(in which one of the two exchanged photons is soft). Thus, terms in which both photons are hard
(and which depend on the hadronic structure) have been ignored.

The most stringent tests of TPE on the nucleon have been carried out by measuring the
ratio of electron and positron elastic scattering off a proton. Corrections due to TPE
will have a different sign in these two reactions. Unfortunately, this (e$^+$e$^-$) data set is 
quite limited\cite{arrington2}, only extending (with poor statistics) up to a \Q-value of $\sim 5$ \GeV, whereas at \Q-values larger than $\sim 2$ \GeV\ basically all data have been measured at 
$\epsilon$-values larger than $\sim 0.85$.

Several studies have provided estimates of the size
of the $\epsilon$-dependent corrections necessary to resolve the discrepancy. 
Because the fall-off
of the form-factor ratio is linear with \Q, and the Rosenbluth formula also shows
a linear dependence of the form-factor ratio (squared) with \Q\ through the $\tau$-term,
a \Q-independent correction linear in $\epsilon$ would cancel the 
disagreement. An additional constraint that any $\epsilon$-dependent modification
must satisfy, is the (e$^+$e$^-$) data set. 

Blunden {\it et al.}\cite{blunden} carried out the first calculation of the elastic contribution from 
TPE effects, albeit with a simple monopole \Q-dependence of the hadronic  form
factors. They obtained a practically 
\Q-independent correction factor with a linear $\epsilon$-dependence that vanishes
at forward angles ($\epsilon = 1$). However, the size of the correction only resolves about
half of the discrepancy. A later calculation which used a more realistic form factor behavior, resolved up to 60\% of the discrepancy.
A different approach was used by Chen {\it et al.}\cite{afanasev}, who related the elastic
electron-nucleon scattering to the scattering off a parton in a nucleon through generalized
parton distributions. TPE effects in the lepton-quark scattering process
are calculated in the hard-scattering amplitudes. The results for the
TPE contribution reduce the discrepancy between the Rosenbluth and the polarization-transfer data by over 50\%. It is highly likely that a combination of the calculations by Blunden {\it et al.} and by Chen {\it et al.}, in which double counting is avoided, could fully reconcile the Rosenbluth and the polarization-transfer data.
Of course, further effort is needed to investigate the 
model-dependence of the TPE calculations. 

Experimental confirmation of TPE effects will be
difficult, but certainly should be continued. The most direct test would be a measurement
of the positron-proton and electron-proton scattering cross-section ratio at small
$\epsilon$-values and \Q-values above 2 \GeV. Positron beams available at
storage rings are too low in either energy or intensity, but a measurement in the CLAS detector at Jefferson Lab, a more promising venue, has been approved\cite{broo2}. A similar measurement, albeit at more limited kinematics, is being prepared at the VEPP-3 storage ring\cite{arring}.
Additional efforts should be extended to studies of TPE effects in other 
longitudinal-transverse separations, such as proton knock-out and 
deep-inelastic scattering (DIS) experiments.

\section{Neutron Magnetic Form Factor}

A significant break-through 
was made by measuring the ratio of quasi-elastic neutron and 
proton knock-out from a deuterium target. This method has little  
sensitivity to nuclear binding effects and to fluctuations in the 
luminosity and detector acceptance. A study of \GMn\ at \Q-values up to 5 \GeV\ has been completed in Hall B 
by measuring the neutron/proton quasi-elastic cross-section ratio using the CLAS detector\cite{broo}.
A hydrogen target was in the beam simultaneously with the deuterium target.
This made it possible to measure the neutron detection efficiency by tagging neutrons in 
exclusive reactions on the hydrogen target. Preliminary results\cite{broo} indicate that
\GMn\ is within 10\% of \GD\ over the full \Q-range of the experiment (0.5-4.8 \GeV), as shown by the red data points in fig. \ref{Chiral}.

Inclusive quasi-elastic scattering of polarized electrons 
off a polarized $^3$He target offers an alternative method to determine \GMn\ through a
measurement of the beam asymmetry\cite{donn}.
By orienting the target polarization parallel to $\vec{q}$,
one measures $R_{T\prime}$, which in quasi-elastic kinematics is dominantly sensitive to $(\GMn)^2$.
For the extraction of \GMn\ corrections for the nuclear medium\cite{gola} are necessary to take into account effects of final-state interactions and meson-exchange currents.

\section{Neutron Electric Form Factor}

In the past decade a series of double-polarization measurements of
neutron knock-out from a polarized $^2$H or $^3$He target have 
provided accurate data on \GEn. The ratio of the beam-target 
asymmetry with the target polarization perpendicular and 
parallel to the momentum transfer is directly proportional to 
the ratio of the electric and magnetic form factors.
A similar result is obtained with an unpolarized deuteron target when one 
measures the polarization of the knocked-out neutron as a function of the angle over which the neutron spin is precessed with a dipole magnet.

 At low \Q-values 
corrections for nuclear medium and rescattering effects can 
be sizeable: 65\% for \deut\ at 0.15 \GeV\ and 50\% for 
\he\ at 0.35 \GeV. These corrections are expected to 
decrease significantly with increasing $Q$. The latest data from Hall C at Jefferson Lab, using either a 
polarimeter \cite{made} or a polarized
 target \cite{warr},  extend up to \Q\ $\approx$ 1.5 \GeV\ with an overall 
accuracy of $\sim$10\%, in mutual agreement. From $\sim$ 1 \GeV\ onwards  \GEn\ appears
to exhibit a \Q-behavior similar to that of \GEp.
Schiavilla and Sick\cite{schi} have extracted \GEn\ from available data on the deuteron quadrupole
form factor $F_{C2}(Q^2)$ with a much smaller sensitivity to the nucleon-nucleon potential
than the extraction from inclusive (quasi-)elastic scattering. 
Wojtsekhowski {\it et al.}\cite{wojt} have recently 
measured \GEn\ in Hall A at four \Q-values between 1.4 and 3.4 \GeV\ using the \Heen\ reaction with a 
100 msr electron  spectrometer, a dedicated 80 ton neutron plastic scintillator detector and a novel polarized $^3$He target using hybrid optical pumping. 
At the BLAST facility  a polarized deuterium target internal to a storage ring has been used successfully to  provide accurate 
data on \GEn\ in a \Q-range from 0.1 to 0.6 \GeV \cite{ziskin}.

\section{Model Calculations}
\label{calc}

\begin{figure}
\resizebox{0.9\textwidth}{!}{%
  \includegraphics{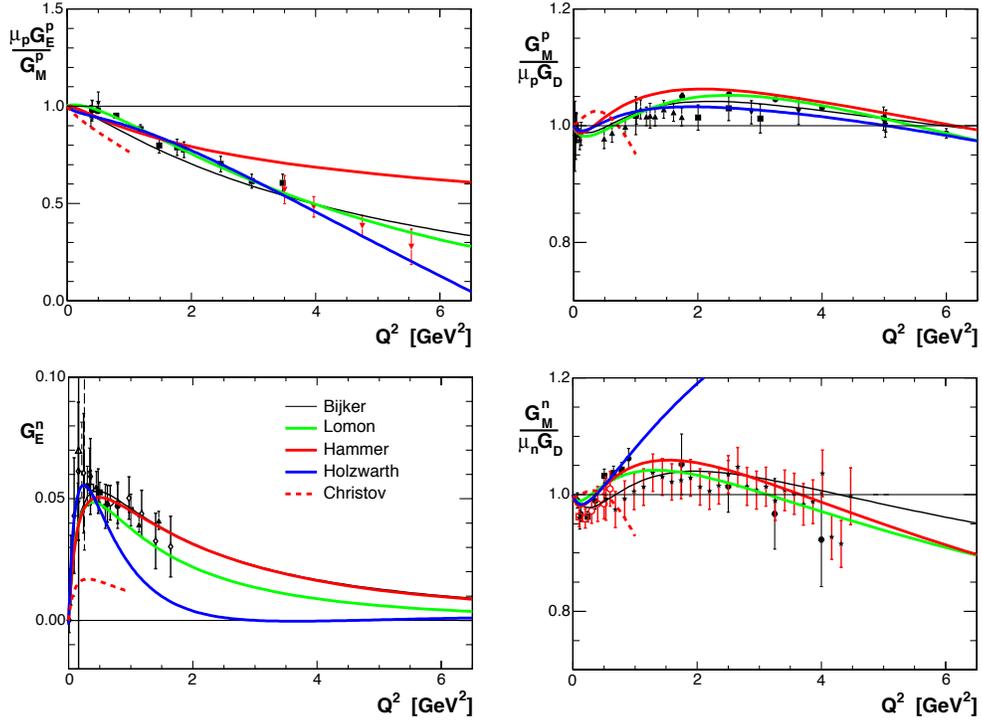}
}
\caption{Comparison of various calculations with available EMFF data. For \GEp\ only polarization-transfer data are shown. For \GEn\ the results of Schiavilla and Sick\cite{schi} have been added.
The calculations shown are from References \cite{iach3,lomo,merg,holz,goek}. Where applicable, the calculations have been normalized to the calculated values of $\mu_{p,n}$. See text for references to the data}
\label{Chiral}       
\end{figure}

The recent production of very accurate EMFF data, especially the surprising \GEp\ data from polarization transfer, has prompted the theoretical community to intensify their investigation of nucleon structure. One expects the three lightest vector mesons ($\rho$, $\omega$ and $\phi$) to play an important role
in the interaction of the photon with a nucleon. The first EMFF models were based on this principle, called vector meson dominance (VMD), in which one assumes that the virtual photon 
- after becoming a quark-antiquark pair - couples to the nucleon 
as a vector meson. With this model Iachello {\it et al.}\cite{iach} predicted a linear drop of the proton form-factor ratio, similar to that measured by polarization transfer, more than 20 years before the data became available. Gari and Kr\"{u}mpelmann\cite{gari} extended the VMD model to conform with pQCD scaling at large \Q-values. The VMD picture is not complete, as becomes obvious from the fact that the
Pauli isovector form factor $F_2^V$ is much larger than the isoscalar one $F_2^S$. An improved description requires the inclusion of the isovector $\pi \pi$ channel through dispersion relations\cite{hohl,merg}. By adding more parameters, such as the width of the $\rho$-meson and the masses of heavier vector mesons\cite{lomo}, the VMD models succeeded in describing new EMFF data as they became available, but with little predictive power. Figure \ref{Chiral} confirms that Lomon's calculations provide an excellent description of all EMFF data. Bijker and Iachello\cite{iach3} have extended the original calculations by also including a meson-cloud contribution in $F_2$, but still taking only two isoscalar and one isovector poles into account. The intrinsic structure of the nucleon is estimated to have an rms radius of $\sim$ 0.34 fm. These new calculations are in good agreement with the proton form-factor data, but do rather poorly for the neutron. 

Many recent theoretical studies of the EMFFs have applied various forms of a relativistic constituent quark
model (RCQM). Nucleons are assumed to be composed of three constituent quarks, which are quasi-particles where all degrees of freedom associated with the gluons and $q \bar{q}$ pairs are parametrized by an effective mass. Because the momentum transfer can be several times the nucleon mass, the constituent quarks require a relativistic quantum mechanical treatment.  Although most of these calculations correctly describe the EMFF behaviour at large \Q-values, effective degrees of freedom, such as a pion cloud and/or a finite size of the constituent quarks, are introduced to correctly describe the behaviour at lower \Q-values.

\begin{figure}[h]
\resizebox{0.9\textwidth}{!}{%
  \includegraphics{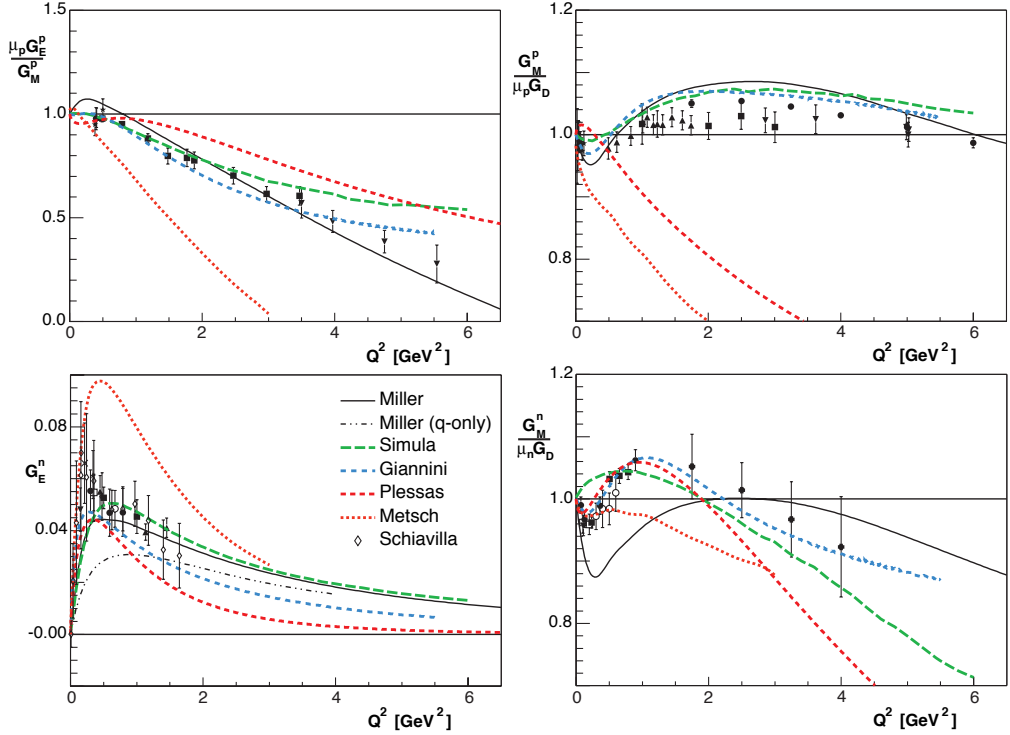}
}
\caption{Comparison of various RCQM calculations with available EMFF data, similar to the comparison in fig. \ref{Chiral}.
The calculations shown are from References \cite{mill1,simu,gian,wage,mets}. Miller (q-only) denotes a calculation by Miller\cite{mill1} in which the pion cloud has been suppressed. Where applicable, the calculations have been normalized to the calculated values of $\mu_{p,n}$. See text for references to the data. For \GEn\ the results of Schiavilla and Sick\cite{schi} have been added.}
\label{RCQM}       
\end{figure}

Miller\cite{mill1} uses an extension of the cloudy bag model\cite{theb}, with three relativistically moving (in light-front kinematics) constituent quarks, surrounded by a pion cloud. Cardarelli and Simula\cite{simu} also use light-front kinematics, but they calculate the nucleon wave function by solving the three-quark Hamiltonian in the Isgur-Capstick one-gluon-exchange potential. In order to get good agreement with the EMFF data they
introduce a finite size of the constituent quarks in agreement with recent DIS data.
The results of Wagenbrunn {\it et al.}\cite{wage} are calculated in a covariant manner
in the point-form spectator approximation (PFSA). In addition to a linear confinement, the quark-quark interaction is based on Goldstone-boson exchange dynamics. The PFSA current is effectively a three-body operator (in the case of the nucleon as a three-quark system) because of its relativistic nature. It is still incomplete but it leads to surprisingly good results for the electric radii and magnetic moments of the other light and strange baryon ground states beyond the nucleon. 
Giannini {\it et al.}\cite{gian} have explicitly introduced a three-quark interaction in the form of a gluon-gluon interaction in a hypercentral model, which successfully describes various static baryon properties. Relativistic effects are included by boosting the three quark states to the Breit frame and by introducing a relativistic quark current.
All previously described RCQM calculations used a non-relativistic treatment of the quark dynamics, supplemented by a relativistic calculation of the electromagnetic current matrix elements. Merten {\it et al.}\cite{mets} have solved the Bethe-Salpeter equation with instantaneous forces, inherently respecting relativistic covariance. In addition to a linear confinement potential, they used an effective flavor-dependent two-body interaction. For static properties this approach yields results\cite{caut} similar to those obtained by Wagenbrunn {\it et al.}\cite{wage}.
The results of these five calculations are compared to the EMFF data in Figure \ref{RCQM}. The calculations of Miller do well for all EMFFs, except for \GMn\ at low \Q-values. Those of Cardarelli and Simula, Giannini {\it et al.} and Wagenbrunn {\it et al.} are in reasonable agreement with the data, except for that of Wagenbrunn {\it et al.} for \GMp, while the results of Merten {\it et al.} provide the poorest description of the data.

Before the Jefferson Lab polarization transfer data on $G_E^p/G_M^p$ became available Holzwarth \cite{holz} predicted a linear drop in a chiral soliton model. In such a model the quarks are bound in a nucleon by their interaction with chiral fields. In the bare version quarks are eliminated and the nucleon becomes a skyrmion with a spatial extension, but the Skyrme model provided an inadequate description of the EMFF data. Holzwarth's extension introduced one vector-meson propagator for both isospin channnels in the Lagrangian and a relativistic boost to the Breit frame. His later calculations used separate isovector and isoscalar vector-meson form factors. He obtained excellent agreement for the proton data, but only a reasonable description of the neutron data. Christov {\it et al.}\cite{goek} used an SU(3) Nambu-Jona-Lasinio Lagrangian, an effective theory that incorporates spontaneous chiral symmetry breaking. This procedure is comparable to the inclusion of vector mesons into the Skyrme model, but it involves many fewer free parameters (which are fitted to the masses and decay constants of pions and kaons). The calculations are limited to \Q $\le$ 1 \GeV\ because the model is restricted to Goldstone bosons and because higher-order terms, such as recoil corrections, are neglected. A constituent quark mass of 420 MeV  provided a reasonable description of the EMFF data (see fig. \ref{Chiral}).

\begin{figure}[h]
\resizebox{0.85\textwidth}{!}{%
  \includegraphics{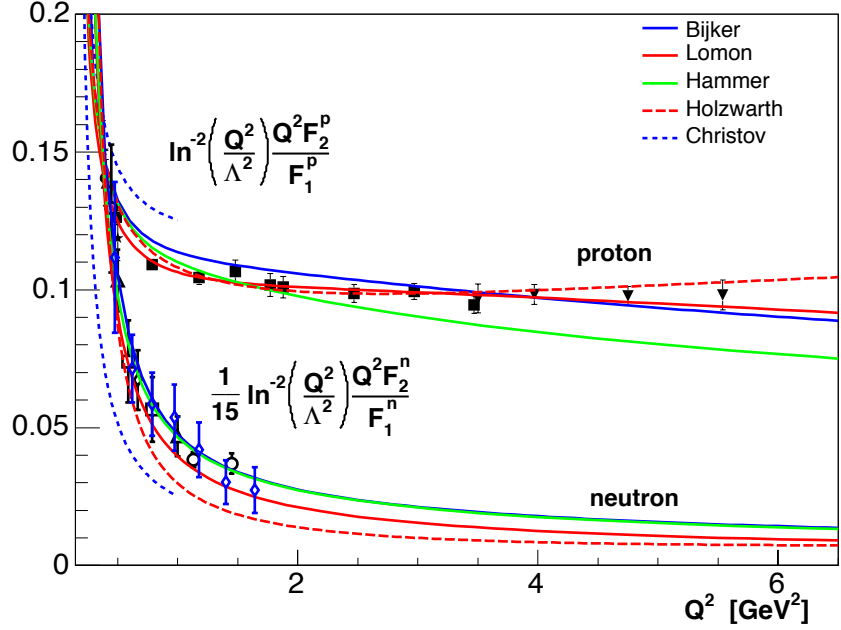}
}
\caption{Comparison of the logarithmic scaling prediction\cite{beli}, assuming $\Lambda$ = 300 MeV, for the proton and the neutron with the available data and a selection of calculations\cite{iach3,lomo,merg,holz,goek}. See text for references to the data.}
\label{lnq2}       
\end{figure}

In the asymptotically free limit, QCD can be solved perturbatively, providing predictions for the EMFF behavior at large \Q-values. Recently, Brodsky {\it et al.}\cite{brod3} and Belitsky {\it et al.}\cite{beli} have independently revisited the pQCD domain. Belitsky {\it et al.} derived the following large \Q-behavior:

\begin{equation} 
\frac{F_2}{F_1} \propto \frac{\ln^2{Q^2/ \Lambda ^2}}{Q^2},
\end{equation}

\setlength{\parindent}{0em}
where $\Lambda$ is a soft scale related to the size of the nucleon.  Figure \ref{lnq2} shows that the polarization-transfer data for the proton appear to follow this behavior already from $\sim$ 1 \GeV\ onwards, as well as the \GEn\ data. However, Belitsky {\it et al.} warn that this could very well be precocious, since pQCD is not expected to be valid at such low \Q-values. In addition Arrington {\it et al.}\cite{arrob} point out that the value of $\sim$ 300 MeV used for  $\Lambda$ corresponds to a length scale of $\sim$ 1 fm which is larger than the nucleon radius.

However, all theories described until now are at least to 
some extent effective (or parametrizations). They use models constructed to focus on certain selected aspects of QCD. Only lattice gauge theory can provide a truly ab initio calculation, but accurate lattice QCD 
results for the EMFFs are still several years away. One of the most advanced lattice calculations of EMFFs has been performed by the QCDSF collaboration\cite{gock}. The technical state of the art limited these calculations to the quenched approximation (in which sea-quark contributions are neglected), to a box size of 1.6 fm and to a pion mass of $\sim$ 500 MeV. Ashley {\it et al.}\cite{ashl} have extrapolated the results of these calculations to the chiral limit, using chiral coefficients appropriate to full QCD. The agreement with the data  is poorer  than that of any of the phenomenological calculations. In a more recent calculation\cite{Alex} the isovector nucleon form factors were calculated both in the quenched approximation and using unquenched configurations for pion masses down to 380 MeV. Although unquenching effects were shown to be small, both quenched and unquenched results are larger than the experimental data. Moreover, the smaller radii obtained showed that pion cloud contributions are underestimated at the pion masses used. Clearly, significant technology developments are required before lattice QCD calculations can provide a stringent test of experimental EMFF data.

\section{The Pion Cloud}

The charge and magnetization rms radii are related to the slope of the form factor at \Q = 0.  Table \ref{radii} lists the results. For an accurate extraction of the radius Sick\cite{sick}
 has shown that it is necessary to take into account Coulomb distortion effects and higher moments of the
radial distribution. His result for the proton charge radius is in excellent agreement with the most recent three-loop QED calculation\cite{meln} of the hydrogen Lamb shift. Within error bars the rms radii for the proton 
charge and magnetization distribution and for the neutron magnetization distribution are
equal. The Foldy term 
$\frac{3}{2} \frac{\kappa}{M_n^2} = -0.126$ fm$^2$ is close to the value of the neutron charge radius.
Isgur\cite{isgur} showed that the Foldy term is canceled by a first-order relativistic correction, which implies that the measured value of the neutron charge radius is indeed dominated by its internal structure. Its negative value can be interpreted as supportive evidence for the picture of a neutron in part behaving as a proton surrounded by a negative pion.

\begin{table}[h!]
\caption{Values for the nucleon charge and magnetization radii}
\label{radii}       
\begin{tabular}{lll}
\hline\noalign{\smallskip}
\textbf{Observable} & \textbf{value $\pm$ error} & \textbf{Reference}  \\
\noalign{\smallskip}\hline\noalign{\smallskip}
$<(r_E^p)^2>^{1/2}$ & 0.895 $\pm$ 0.018 fm & \cite{sick} \\
$<(r_M^p)^2>^{1/2}$ & 0.855 $\pm$ 0.035 fm & \cite{sick} \\
$<(r_E^n)^2>$ & - 0.119 $\pm$ 0.003 fm$^2$ & \cite{kope} \\
$<(r_M^n)^2>^{1/2}$ & 0.87 $\pm$ 0.01 fm & \cite{kubo} \\
\noalign{\smallskip}\hline
\end{tabular}
\end{table}

\begin{figure}[h!]
\resizebox{0.9\textwidth}{!}{%
  \includegraphics{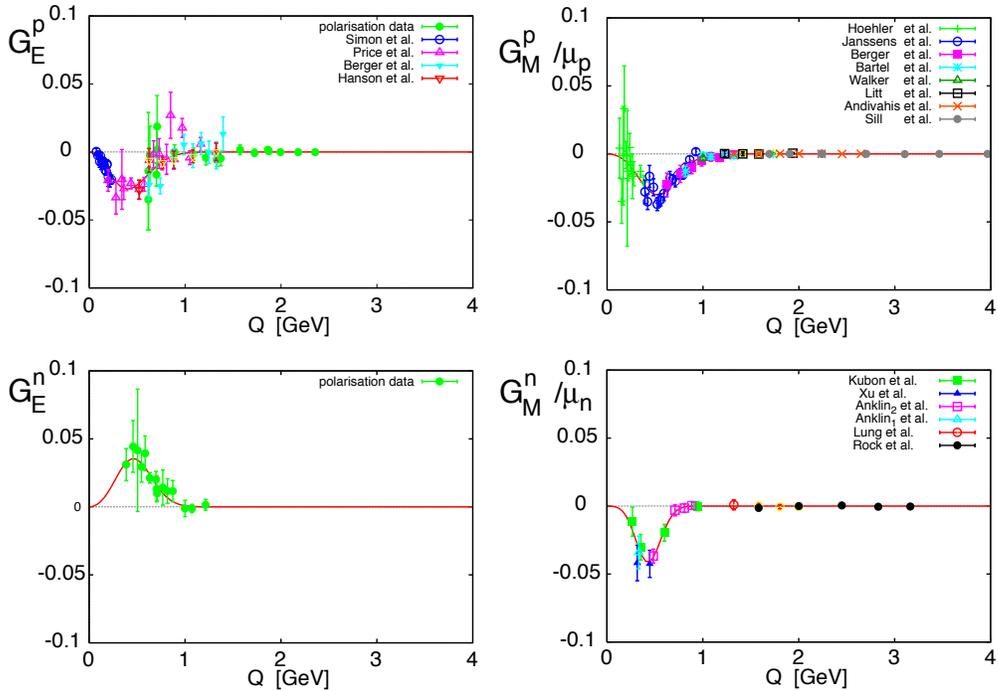}
}
\caption{Available data for the four EMFF, plotted as a function of $Q$. All four form factors show an indication of structure at $\sim$ 0.5 GeV. The figure has been adapted from ref. \cite{fried}, where also the references to the data can be found.}
\label{FWq2}       
\end{figure}

In the Breit frame the nucleon form factors can be written as Fourier transforms of their charge
and magnetization distributions. However, if the wavelength of the probe is larger than the Compton wavelength of the nucleon, i.e. if $| Q | \ge M_N$, the form factors are not solely determined by the internal structure of the nucleon. Then, they also contain dynamical effects due to relativistic boosts and consequently the physical interpretation of the form factors becomes complicated. Kelly\cite{kelly} has extracted spatial nucleon densities from the 
available form factor data. He selected a model for the Lorentz contraction of the Breit frame in which the asymptotic behavior of the form factors conformed to perturbative quantum chromo-dynamics (pQCD) scaling at large \Q-values and expanded the densities in a complete set of radial
basis functions, with constraints at large radii. The neutron and proton magnetization densities
are found to be quite similar, narrower than the proton charge density. He reports a neutron charge density with a positive core surrounded by a negative surface charge, peaking at just below 1 fm, which he attributes to a negative pion cloud. Friedrich \& Walcher\cite{fried} observe as a feature common to all EMFFs a bump/dip at $Q \approx$ 0.5 GeV with a width of $\sim$ 0.2 GeV, as is illustrated in fig. \ref{FWq2}. 
A fit to all four EMFFs was performed, assuming a dipole behaviour for the form factors of the constituent quarks and an $l = 1$ harmonic oscillator behaviour for that of the pion cloud. They then transformed their results to coordinate space, neglecting the Lorentz boost, where they find that the pion cloud peaks at a radius of $\sim$ 1.3 fm, slightly larger than Kelly did, close to the Compton wavelength of the pion. 

Hammer {\it et al.}\cite{hamm2} argue from general principles that the pion cloud should peak much more inside the nucleon, at $\sim$ 0.3 fm. However, they assign the full $N\bar{N}$2$\pi$ continuum to the pion cloud which includes different contributions than just the one-pion loop that Kelly (and Friedrich \& Walcher) assign to the pion cloud. The structure at $\sim$ 0.5 GeV, common to all EMFFs, is at such a small \Q-value that its transformation to coordinate space should be straightforward.

Several theoretical models for the nucleon have explicitly included, albeit phenomenologically, the effect of a pionic cloud. Miller\cite{mill1} has shown that in his cloudy-bag model the pionic contributions to \GEn\ dominate at small \Q\-values (see fig. \ref{RCQM}). Faessler {\it et al.}\cite{faessler} have developed a Lorentz covariant chiral quark model, in which nucleons are considered bound states of constituent quarks further dressed by a cloud of pseudoscalar mesons. In a first step the parameters of their chirally symmetric Lagrangian are fitted to the magnetic moments of the baryonic octet. Next, the form-factor data for \Q\ $\ge$ 0.7 \GeV\ are fitted with a dipole form multiplied by a gaussian, with a total of ten free parameters. Then, the pseudoscalar meson cloud contribution is fixed through chiral perturbation theory on the hadron level. The meson cloud contribution is forced through a gradual cut-off function to be strongly suppressed for large \Q\ -values. The results provide an excellent description of the available form-factor data. Interestingly, the bump (dip) at $\sim$ 0.5 GeV in \GEn\ (\GEp\ ) is attributed completely to the meson cloud, whereas the meson cloud already contributes to the static magnetic moment with a \Q\ -behaviour peaking at \Q = 0 \GeV. Bhagwat {\it et al.}\cite{bhag} have applied the Dyson-Schwinger equations to calculate the nucleon electro-magnetic form factors. They predict that \GEpGMp\ will pass through zero at \Q $\approx$ 6.5 \GeV. Their calculations show that indeed the small \Q\ -behaviour of \GEn\ is dominated by the neutron's pion cloud, whereas the evolution of \GEn\ for \Q $\ge$ 2 \GeV\ will primarily be determined by the quark-core of the neutron. Thus, they predict that the ratio of \GEn\ and \GMn\ will continue to increase steadily until \Q $\approx$ 8 \GeV.

\section{Experimental Review and Outlook}

In recent years highly accurate data on the nucleon EMFFs have become available from various facilities
around the world, made possible by the development of high luminosity and novel polarization
techniques. These have established some general trends in the \Q-behavior of the four EMFFs. 
The two magnetic form factors \GMp\ and \GMn\ are close to identical, following \GD\ to within 10\%
at least up to 5 \GeV, with a shallow minimum at $\sim 0.25$ \GeV\ and crossing \GD\ at $\sim 0.7$ \GeV.
\GEpGMp\ drops linearly with \Q\, and \GEn\ appears to drop from $\sim 1$ \GeV\ onwards at the same rate as \GEp.
Highly accurate measurements with the Rosenbluth technique have established that the discrepancy between results on \GEpGMp\ with the Rosenbluth techniques and with polarization transfer is not an instrumentation problem. Recent advances on two-photon exchange contributions make it highly likely that the application of TPE corrections will resolve that discrepancy. 

There remains a strong disagreement about the interpretation of the apparent structure in the four nucleon form factors at $\sim$ 0.5 GeV. While a seemingly straightforward picture of a pionic cloud surrounding the nucleon can explain the structure, a recent dispersion analysis\cite{belu} indicates that the $2\pi$ continuum contributions are much more confined in coordinate space. The structure in the data can only be reproduced by additional low-mass strength in the spectral function in a $\Q$-region that is already well understood. Thus, there is a clear need for additional data of very high accuracy in the $Q$-range from 0.3 to 0.8 GeV.

Measurements that extend to higher \Q-values and offer improved accuracy at
lower \Q-values, will become available in the near future. In Hall C at Jefferson Lab Perdrisat {\it et al.}\cite{perd} 
will extend the measurements of \GEpGMp\ to 9 \GeV\ with a new polarimeter and 
large-acceptance lead-glass calorimeter. In Hall A a proposal\cite{gilman} has been submitted to measure \GEpGMp\ using the recoil polarimeter technique with an accuracy approaching 1\% in a \Q-range between 0.2 and 0.8 \GeV.
Once the upgrade to 12 GeV\cite{12gev}  has been implemented
at Jefferson Lab, it will be possible to extend the data set on \GEp\ and \GMn\ to
15 \GeV\ and on \GEn\ to 8 \GeV.

\section*{Acknowledgments}

This work was supported by DOE contract DE-AC05-84ER40150 Modification No. M175, under which the Southeastern Universities Research Association (SURA) operates the Thomas Jefferson National Accelerator Facility. 


\end{document}